# Heavily hydride-ion-doped 1111-type iron-based superconductors: Synthesis, physical properties and electronic structure

Soshi Iimura[1, 2] and Hideo Hosono[1]

[1]Materials Research Center for Element Strategy, Tokyo Institute of Technology, 4259 Nagatsuta, Midori-ku, Yokohama 226-8503, Japan

[2]PRESTO, Japan Science and Technology Agency, 4-1-8 Honcho, Kawaguchi 332-0012, Japan

[*] H. Hosono, e-mail: hosono@mces.titech.ac.jp

# Abstract

Iron-based superconductors have grown to be a new continent of high $T_c$ superconductors comparable to cuprates. The optimal critical temperature ($T_c$) of 56K in electron-doped 1111-type iron oxypnictides attracts considerable attention of physicists and chemists. Carrier doping is not only essential to induce superconductivity but also is a critical parameter that governs the electronic, magnetic, and crystallographic properties of ground states in high-$T_c$ superconductors. Hydride ion ($H^-$) which is an anionic state of hydrogen acts as an efficient electron donor in the 1111-type, leading to several important discoveries such as two-dome structure of superconducting phase and bipartite parent phase. This article summarizes the synthesis, physical properties, and electronic structure of the $H^-$ bearing 1111-type iron-based superconductors along with the relevant phenomena of other superconductors. We show several common characteristics of iron-based superconductors with high-$T_c$ over 50 K and suggest a way to achieve higher $T_c$.

# 1. Introduction

In 2006, more than 10 years passed since the very rapid growth of $T_c$ in cuprates was leveled off, Kamihara *et al*. reported a first iron phosphide superconductor LaFePO with $T_c = 4K$[1]. In 2008, $T_c$ rose to 26 K in iron arsenide LaFeAsO$_{1-x}$F$_x$[2]. Within two months from this report, $T_c$ of LaFeAsO$_{1-x}$F$_x$ jumped to 43 K by applying a pressure of 4GPa, which exceeds $T_c$ of MgB$_2$ (39K)[3]. Now the optimum $T_c$ has increased to 55-58K by replacing La with Sm[3-9], qualifying the iron-based superconductors (IBSCs) as the second family of high-$T_c$ superconductors. The series of discoveries sparked overwhelming research activities directed to explore much higher $T_c$ in the iron-based compounds.

**Fig. 1** shows a summary of crystal structures of various IBSCs. The IBSCs share a common [Fe*Pn*]$^-$ (*Pn* = pnictogen) or [Fe*Ch*] (*Ch* = chalcogen) layer composed of the edge-sharing Fe*Pn*$_4$ or Fe*Ch*$_4$ tetrahedra (**Fig. 1(a)**)[10]. The nearest neighbor Fe-Fe distance is ~ 2.6-2.8 Å, while the Fe-*Pn* or Fe-*Cn* is ~ 2.3-2.6 Å. *A*Fe*Pn* (*A* = alkali metal) or *AE*Fe$_2$*Pn*$_2$ (*AE* =alkaline earth metal), abbreviated respectively as 111 or 122-type, is the simplest structure of iron pnictides (**Fig.s 1(b) and (c)**)[11,12]. In the 111-type, the alkali metals are located inside the *Pn*$_5$ pyramid, while in the 122-type, the larger alkaline earth metals are positioned inside the *Pn*$_8$ rectangle. The 111-type crystallizes in a primitive tetragonal (*P4/nmm*), while 122-type does body-centered tetragonal (*I4/mmm*) due to the sliding of FeAs layer each other. More complex structures are obtained by inserting a 2-dimensional square net, skutterudite-type layer, zig-zag chain, and perovskite block into between the *A* ions of the 111-type, abbreviated respectively as 1111 (**Fig. 1(d)**)[1], 1048 (**Fig. 1(e)**)[13], 112 (**Fig. 1(f)**)[14] and 42622-type (**Fig. 1(g)**)[15]. If the Fe*Pn* layers and the *AE* ions of 1048 and 42622 is shifted horizontally against each other, the other derivatives, 1038[13] and 32522[16], respectively, are obtained.

Fe*Ch* (*Ch* = S, Se, Te) abbreviated as 11-type is the simplest form of iron chalcogenides having anti-PbO type structure (**Fig. 1(h)**)[17–19]. Various cations such as $A$ = Li⁺, Na⁺, K⁺, Rb⁺, Cs⁺, $AE$ = Ca²⁺, Sr²⁺, Ba²⁺, and $RE$ = Eu²⁺, Yb²⁺ (*RE* = rare-earth), can be inserted between the FeSe layers, where for the larger $A$ ions such as K⁺, Rb⁺, Cs⁺ (**Fig. 1(i)**)[20,21] a high temperature solid state reaction is required, while for the smaller $A$, $AE$ and $RE$ cations a low temperature intercalation reaction is effective[22]. The intercalation of the small cation sometimes accompanies with solvent molecules such as ammonia (**Fig. 1(i)**)[23], pyridine ($C_5H_5N$)[24], ethylenediamine ($C_2H_8N_2$)[25]. To compensate the charge neutrality, a large amount of vacancies in the Fe site is formed through the incorporation of the cations. As a result, the chemical composition ratio of the intercepted iron chalcogenides approaches to $A^+ : Fe^{2+} : Se^{2-}$ = 2 : 4 : 5 (**Fig. 1(k)**)[21].

Electronic phase diagram that plots transition temperatures as a function of carrier concentration or applied pressure is a useful tool to overview the physical properties of superconductors. In **Fig.s 2(a) and (b)** the electronic phase diagrams of LaFeAsO₁₋ₓFₓ[2,26,27] and Ba(Fe₁₋ₓCoₓ)₂As₂[28,29] are shown, respectively. For 1111-type LaFeAsO, a second order structural (from tetragonal to orthorhombic) transition followed by a magnetic (from paramagnetic (PM)-antiferromagnetic (AFM)) transition is observed on cooling at separated temperature (hereafter we refer the transition temperatures to $T_s$ and $T_N$, respectively)[30,31]. Superconductivity emerges when the orthorhombic AFM phase is completely suppressed by doping electrons via F⁻ substitution for oxygen site (O²⁻ → $F^-$ + $e^-$). For 122-type Ba(Fe$_{1-x}$Co$_x$)$_2$As$_2$, the $T_s$ and $T_N$ coincides each other at the non-doped sample, and thus the structural and magnetic transitions become a first order. The superconducting phase appears before the structural and magnetic transitions disappear, in other words, they coexist microscopically or macroscopically in the slightly doped region. The $T_c$ takes an optimum value at which the structural and magnetic transitions are completely suppressed, implying that quantum critical fluctuations play a key role in the emergence of superconductivity in 122-type[32]. Since the superconductivity of IBSCs emerges when the structural and magnetic transitions are suppressed, the orthorhombic AFM phase is called a parent phase for the superconductivity.

To explain the emergence of dome-shape superconducting phase near the AFM phase, a spin fluctuations model resulting from Fermi surface (FS) nesting between hole and electron pockets was proposed based on density functional theory (DFT) calculations[33,34]. **Fig. 2(c)** is a band structure of tetragonal LaFeAsO. The ten bands around Fermi level are composed mainly of Fe-3$d$ orbitals. Among them, five bands crossing the Fermi level form hole pockets at around Γ point and electron pockets at around M point. Since LaFeAsO crystallizes in a 2-dimensional layered structure, the dispersion along $k_z$ direction is poor, yielding cylindrical FSs at around Γ and M points (see **the inset of Fig. 2(c)**). The cylindrical hole and electron FS with a similar curvature has a strong FS nesting in the direction indicated by the arrow in **the inset of Fig. 2(c)**, and produces a peak of spin susceptibility at the corresponding wave number. Therefore, the AFM is believed to be developed by a spin density wave (SDW) mechanism, and the superconductivity is to be mediated by the fluctuations of the SDW. On the other hand, a heavy electron doping over $x > 0.1$ degrades the FS nesting, because the hole pockets are reduced and the electron pockets are expanded[35,36]. As a result, spin fluctuations are weaken, and thus, the superconductivity disappears by the heavy doping. This spin fluctuations model explains not only dome shape of superconducting phase in IBSCs upon electron doping, but also a large enhancement of $T_c$ from 26K in LaFeAsO$_{1-x}$F$_x$ to the highest $T_c$ of 56K among IBSCs by replacing La with Sm[37].

Contrary to the high $T_c$ of 1111-type compounds, the phase diagrams reported so far for $RE$FeAsO$_{1-x}$F$_x$ were rather incomplete[38–42]. For instance, no experimental data on the over-doped region of $RE$FeAsO$_{1-x}$F$_x$ had been reported until 2011 (except $RE$ = La). This situation primarily comes from the formation of thermodynamically stable $RE$OF phase upon F doping over $x > 0.10$, which restricts the solubility of F in 1111-type[43,44].

## 2. Hydride ion substitution into $RE$FeAsO

### 2.1. Why H⁻ doping ?

To solve the serious issue in the electron doping into 1111-type, we proposed hydride ion (H⁻) as an alternative electron dopant for fluoride ion[45–47]. Hydrogen has a moderate electronegativity of 2.1 and then behaves as a simplest bipolar element in response to the local environment. If highly electro-negative elements such as oxygen and fluorine are bonded to hydrogen, a proton (H⁺) is formed by losing an electron from the 1$s$ orbital (**Fig.s 3(a)**). On the other hand, hydrogen anion, H⁻, with a 1$s^2$ electronic configuration is stabilized by electro-positive elements such as alkali, alkaline earth and rare earth metals. The ionic radius of H⁻ ranging from 1.15 to 1.65Å[48] is much larger than the size of H⁺ ($0.88 \times 10^{-5}$Å)[49] and H⁰ (0.32 Å)[50], rather similar to those of O$^{2-}$ and F⁻ (~1.3Å).

Due to the close ionic radii of $H^-$ and $O^{2-}$, both anions can occupy a crystallographically similar site. **Fig.s 3(b)** show the crystal structure of a first oxyhydride LaHO reported in 1982[51–53]. It crystallizes in a fluorite superstructure, in which both the $O^{2-}$ and $H^-$ ions occupy the center of La4 tetrahedra, as $O^{2-}$ ion does in the $[LaO]^+$ layer of La-1111. Hence, if half of the anions in LaHO is replaced with $[FeAs]^-$ of La-1111, the resulting structure corresponds to the $H^-$-doped LaFeAsO (LaFeAsO$_{0.5}$H$_{0.5}$), where the substituted $H^-$ for the oxygen site acts as an electron donor: $O^{2-} \rightarrow H^- + e^-$.

## 2.2. How to synthesize and control the doped hydride content

The preparation of oxyhydrides requires a special care due to the high dissociation pressure of hydrogen. A topotactic reaction is the one of the widely used methods to synthesize the oxyhydride[54–60]. The powder of host oxide phase is ground with an excess metal hydride such as $CaH_2$. Heating the mixture removes the oxygen via the formation of CaO, and hydride ion occupies the vacant site of oxigen. Since only the oxygen and hydrogen are exchanged and no bond-breaking and -forming occur in the host structure, the topotactic reaction can proceed at low temperature and prevent the desorption of hydrogen gas during synthesis.

However, this method is not suitable for researches on superconductors because it is difficult to control the doping concentration accurately. Another reason is the difficulty in measuring the transport properties of synthesized samples. The resistivity and Hall effect measurement require high density pellets or single crystals to minimize the effects of grain boundary scattering. However, for the topotactic reaction, the powder precursors are highly preferred to increase the interfacial area between the oxide phase and $CaH_2$ to accelerate the reaction rate.

Solid state reaction at high pressure has some advantages to synthesizing hydride phases over a conventional synthesis at ambient atmosphere. One is related to the large volume reduction associated with the formation of hydride or oxyhydride. According to Le Chatelier's principle, chemical reactions at high pressure proceed in a direction that reduces the total volume. The formation of hydride or oxyhydride accompanies a large volume reduction, because the formation of metal hydride proceeds by dissolving gaseous hydrogen into the oxide host. In addition, the slightly smaller ionic radius and more deformable nature of hydride ion compared to oxide ion also act favorably for the formation of hydrides at high pressure. Second, a repulsive inter-$H_2$ molecule interaction of gaseous hydrogen also promotes the formation of hydrides under high pressure[61,62]. **Fig. 4(a)** shows the pressure dependence of chemical potential of gaseous hydrogen. The variation is in line with the tendency of an ideal gas shown as a dotted line up to 1GPa, above which the Gibbs energy increases rapidly. This increase at high pressures is nearly independent of temperature and originates from the increase of the enthalpy due to repulsive interactions between $H_2$ molecules. As a consequence, the solubility of hydrogen in metal is largely enhanced as synthetic pressures increased especially above 1GPa (**Fig. 4(b)**). The last is that high pressure synthesis makes it easy to prepare a dense pellet of hydrides, which is useful for evaluating the transport properties of superconductors.

Polycrystalline of $RE$FeAsO$_{1-x}$H$_x$ is prepared using a mixture of precursors, $RE_2O_3$, $RE$H$_2$, $RE$As, Fe$_2$As, and FeAs, according to the equation[45,63]:

$$(2 + x)RE\text{As} + (2 + x)\text{Fe}_2\text{As} + (2 - 2x)\text{FeAs} + (2 - 2x)RE_2\text{O}_3 + 3xRE\text{H}_3 \rightarrow$$

$$6RE\text{FeAsO}_{1-x}\text{H}_x + 3/2x\text{H}_2 \uparrow \tag{1}$$

**Fig. 4(c)** illustrates the schematic of high pressure cell set-up. The sample cell is mainly composed of 90 wt%-NaCl and 10 wt%-ZrO$_2$. The outside of the cell is covered by pyrophyllite (Al$_2$Si$_4$O$_{10}$(OH)$_2$) gasket as a pressure medium. The graphite sleeve is used as a resistance heater and two Mo discs for electric lead are attached below and above the graphite sleeve. The hexagonal boron nitride (hBN) sleeve of 6 mm (7 mm) in inner (outer) diameter and 8 mm in length is inserted into the NaCl (+10 wt%ZrO$_2$) tube, and the pellet of mixture of precursors is placed inside the hBN sleeve. Two pellets (6 mm in diameter and 2 mm in thickness) mixture of NaBH$_4$ and Ca(OH)$_2$ with a molar ratio of 1:2 was placed below and above the sample pellet to supply excess hydrogen through the following reaction[64,65]:

$$\text{NaBH}_4 + 2\text{Ca(OH)}_2 \rightarrow \text{NaBO}_2 + 4\text{H}_2\uparrow. \tag{2}$$

Finally, the target phase $RE\text{FeAsO}_{1-x}\text{H}_x$ was prepared by heating at 1200°C and 2-5GPa for 30min.

The amount of hydrogen incorporated in 1111-type can be controlled by tuning the nominal amount of oxygen deficiency. In equations (1) and (2), hydrogen much excess to the amount of oxygen deficiency ($x$) is supplied. Unlike very small H$^+$, H$^-$ does not occupy interstitial sites so much due to the large ionic radius. Furthermore, H$^-$ is only stabilized at a specific, highly electropositive site such as the center of $RE$4 tetrahedra in 1111-type structure. Therefore, even if the mixture of precursors is exposed to excess hydrogen gas, we thought that H$^-$ is captured only by the oxygen vacancy site and the remaining hydrogen gas escapes from the system.

**Fig.s 5(a)** and **(b)** show neutron powder diffraction (NPD) patterns of CeFeAsO$_{1-x}$D$_x$ with nominal $x$ ($x_{nom.}$) = 0.3[46]. The pattern in **Fig. 5(a)** is fitted using the deuteride ion-substituted model, while in **Fig. 5(b)** the oxygen-deficient model is used. It is obvious that the deuteride ion-substituted model produces a much better fitting result. The measured atomic ratio of Ce, As, O and H per Fe in CeFeAsO$_{1-x}$H$_x$ is summarized in **Fig. 5(c)**. The amount of hydrogen incorporated in 1111-type linearly increases with $x_{nom.}$, while oxygen amounts decreases. Moreover, the summation of H and O takes almost unity irrespective of $x_{nom.}$. These data indicate that hydrogen successfully substitutes oxygen site, and the hydrogen amount can be precisely controlled by the amount of oxygen vacancy ($x_{nom.}$).

**Fig.s 6(a) and (b)** show the electronic phase diagram of SmFeAsO$_{1-x}$H$_x$ and CeFeAsO$_{1-x}$H$_x$[45,46]. In the figures, the phase transition temperature of F$^-$-doped samples are also plotted[43,66]. The $x$ dependence of $T_c$ in SmFeAsO$_{1-x}$H$_x$ and CeFeAsO$_{1-x}$H$_x$ coincides well with those of F$^-$-doped samples, indicating that both dopant acts as electron donor and supply the same amount of electrons into the FeAs layer. The $T_c$ of SmFeAsO$_{1-x}$H$_x$ at $x$ > 0.15 is robust with respect to the variation of $x$ and slightly decreases at $x$ > 0.3, and finally the superconductivity disappears at $x$ = 0.47. A similar trend is also observed in CeFeAsO$_{1-x}$H$_x$. The optimum $T_c$ of 56K and 48K are obtained for SmFeAsO$_{1-x}$H$_x$ and CeFeAsO$_{1-x}$H$_x$, respectively. The former value is similar to that of SmFeAsO$_{1-x}$F$_x$, while the latter is slightly higher than that of CeFeAsO$_{1-x}$F$_x$ (41K)[5].

### 2.3. Critical effect of atmosphere during synthesis

The controllability of doped $H^-$ content in 1111-type is based on the idea that the amount of hydrogen in 1111-type is determined by the amount of oxygen vacancy, that is, oxygen vacancy is filled with $H^-$ ion. To examine this idea, we performed several high pressure syntheses with special cares[67]; First, we carefully annealed precursors under vacuum to remove unintentionally contained and/or adsorbed $H_2$ and $H_2O$ molecules on them. Second, the sample assembly for high-pressure synthesis are placed in a stainless-steel capsule and are sealed by welding to prevent contamination of $H_2$ and $H_2O$ from the NaCl cell and pyrophyllite. Last, three different atmospheres around the sample are prepared: hydrogen gas (HYD), water (WAT), and arogon without $H_2$ or $H_2O$ (NONE). The $H_2$ or $H_2O$ atmosphere was created in the welded stainless-steel capsule by placing $NH_3BH_3$ or $Ca(OH)_2$ which respectively releases $H_2$ or $H_2O$ through the thermal decomposition reactions. The influence of $H_2$ and $H_2O$ gas on 1111-type is examined by exposing the mixture of precursors with a nominal composition ratio of $Sm : Fe : As : O = 1 : 1 : 1 : 1 - x_{nom.}$ to the atmosphere;

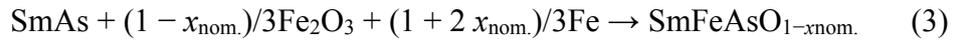

$$SmAs + (1 - x_{nom.})/3Fe_2O_3 + (1 + 2\,x_{nom.})/3Fe \rightarrow SmFeAsO_{1-xnom.} \qquad (3)$$

**Fig. 7(a)** shows the $x_{nom.}$ dependence of weight fraction of the resulting 1111-type. Although the fraction in HYD and WAT cases is almost unity irrespective of $x_{nom.}$, for the NONE case, a large amount of SmAs and Fe impurity phases appear and the fraction linearly decreases with $x_{nom.}$. The lattice parameters of NONE case sample remain unchanged with $x_{nom.}$, while those of HYD and WAT samples monotonically decrease as shown in **Fig. 7(b)**. **Fig.s 7(c)** summarize the $x_{nom.}$ dependence of oxygen and hydrogen amounts in each 1111 phase synthesized under the three conditions. For HYD case, the hydrogen amount increases and the oxygen decreases, both of which are in line with the tendency of nominal amount. For WAT case, the amount of hydrogen is less than that of HYD case. In both the HYD and WAT cases, the summation of hydrogen and oxygen amount is ~1 irrespective of $x_{nom.}$, indicating that the hydrogen from gaseous $H_2$ or $H_2O$ substitutes the oxygen site. On the other hand, the hydrogen in 1111 phase is hardly detected for the NONE case. In **Fig. 7(d) and (e)**, the lattice parameters and $T_c$ of HYD and WAT cases samples are compared with those of $SmFeAsO_{1-x}H_x$ These data are in line with the data of $SmFeAsO_{1-x}H_x$, indicating that the 1111 phase prepared by the HYD and WAT condition are $SmFeAsO_{1-x}H_x$.

These experimental observations can be summarized in the following reactions:

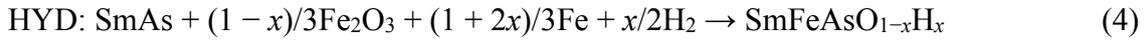

HYD: $SmAs + (1 - x)/3Fe_2O_3 + (1 + 2x)/3Fe + x/2H_2 \rightarrow SmFeAsO_{1-x}H_x$     (4)

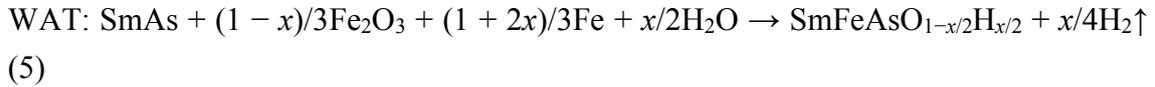

WAT: $SmAs + (1 - x)/3Fe_2O_3 + (1 + 2x)/3Fe + x/2H_2O \rightarrow SmFeAsO_{1-x/2}H_{x/2} + x/4H_2\uparrow$
(5)

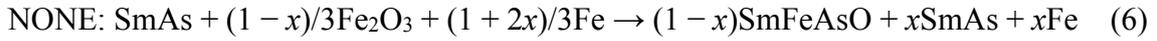

NONE: $SmAs + (1 - x)/3Fe_2O_3 + (1 + 2x)/3Fe \rightarrow (1 - x)SmFeAsO + xSmAs + xFe$     (6)

The important findings through this experiments are as follows; (1) If both the oxygen and hydrogen are present, oxygen preferentially occupies the oxygen vacancy site as $O^{2-}$. (2) After all oxygen is taken into the oxygen vacancy of 1111 phase, hydrogen is incorporated into the remaining oxygen vacancy site as $H^-$. (3) The non-doped SmFeAsO as well as secondary phases such as SmAs and Fe are formed, if neither hydrogen, oxygen nor water are supplied to the oxygen-deficient 1111 phase. In other word, 1111-type does not prefer to form a vacancy at oxygen site.

## 3. Physical properties

### 3.1. Two domes in $T_c$-$x$ curve

Since the heavy electron doping over $x = 0.2$ enlarges the electron pocket at around M point and shrinkages the hole pocket at $\Gamma$ point, this mismatch in the size of electron and hole pockets deteriorates the FS nesting, and as a consequence, the spin fluctuations and superconductivity are suppressed[33–36]. This spin fluctuations model seems to be consistent with the phase diagrams of LaFeAsO$_{1-x}$F$_x$, 122-type and 111-type[68] whose $T_c$ are suppressed at $x$ less than 0.2 (see **Fig. 2(a) and (b)**). However, for SmFeAsO$_{1-x}$H$_x$ and CeFeAsO$_{1-x}$H$_x$, their $T_c$ rather increases up to $x = 0.2$, takes a maximum value, and is finally suppressed by a much heavier electron doping level over 0.4, throwing a serious question into the spin fluctuations model (see **Fig. 6(a) and (b)**). In addition, a basic question is also raised: why the superconducting dome width and $T_c$ are so different between La-1111 and (Ce, Sm)-1111. For instance, the temperature dependence of resistivity in the normal conducting state indicates that the SmFeAsO$_{1-x}$H$_x$ and CeFeAsO$_{1-x}$H$_x$ behave non-Fermi liquid, that is, $\rho(T) \sim T$, whereas LaFeAsO$_{1-x}$F$_x$ obeys Fermi liquid, that is, $\rho(T){\sim}T^2$. These differences imply that the electron doping via fluorine substitution in LaFeAsO$_{1-x}$F$_x$ is insufficient to draw out the genuine physical property of LaFeAsO.

Thus, we synthesized H$^-$-doped LaFeAsO and examined the superconducting properties. The $T_c$-$x$ curve for H$^-$-doped LaFeAsO is shown in **Fig.s 8(a)**[27,47]. For a lightly doped region with $x$ less than 0.2, the $T_c$ of LaFeAsO$_{1-x}$H$_x$ and LaFeAsO$_{1-x}$F$_x$ coincide each other, and form a dome-shaped superconducting dome in the range of $x = 0.04$-$0.20$ (SC1). However, a further electron doping over $x = 0.2$ re-raises $T_c$ up to 33K (for resistivity measurement the optimum $T_c$ is 36K) at $x = 0.36$ and forms a second superconducting dome in the $x$ ranging 0.20-0.45 (SC2). The temperature dependence of resistivity above the $T_c$ of SC1 and SC2 is totally different (**Fig. 8(b)**): the former behaves as a Fermi liquid, while the later does as a non-Fermi liquid in which the resistivity increases linearly with temperature[69]. By replacing the La site with smaller *RE* elements such as Ce, Sm, Gd, and Dy, the two domes of LaFeAsO$_{1-x}$H$_x$ are merged into a single dome, where the optimum $T_c$ of SmFeAsO$_{1-x}$H$_x$ or GdFeAsO$_{1-x}$H$_x$ reaches to 54K (**Fig. 8(c)**).

The sensitivity of $T_c$-$x$ curve shape to the crystal structure can also be seen when pressure is applied to 1111-type[47,70]. **Fig. 8(d)** is the $T_c$-$x$ curves of LaFeAsO$_{1-x}$H$_x$ or SmFeAsO$_{1-x}$H$_x$ at different pressures. The $T_c$ = 18K at the valley of the two-dome structure steeply increases to 44K by applying 3GPa, yielding a single dome structure as like a CeFeAsO$_{1-x}$H$_x$. The optimum $T_c$ of 52K is attained at 6GPa, above which the $T_c$ is decreased to 45K at 13GPa. Contrary to the LaFeAsO$_{1-x}$H$_x$, applying pressure into SmFeAsO$_{1-x}$H$_x$ monotonically decreases the $T_c$.

### 3.2. Two-dome structure in other system

A phosphorus substitution for the arsenic site of 1111-type is another way to tune the superconducting properties[71]. **Fig. 9(a)** shows the electronic phase diagram of SmFeAs$_{1-y}$P$_y$O$_{1-x}$H$_x$. The P substitution monotonically reduces the $T_c$ irrespective of H contents ($x$), and changes the shape of $T_c$-$x$ curve from the high $T_c$ single-dome to the two-dome structure with a $T_c$ valley at $x \sim 0.2$. These results indicate that if a certain structural condition is satisfied, the two-dome structure appears regardless of the type of *RE*. A similar two-dome structure is also observed in other iron phospho-arsenides such as $Ca_4Al_2O_6Fe_2(As_{1-x}P_x)_2$[72] and $LaFeAs_{1-x}P_xO$[73].

The two-dome structure also emerges in iron chalcogenides[74–78]. $A_x(NH_3)_yFe_2Se_2$ is an electron-doped iron selenide system, and the electron is doped through intercalation of *A* ions occurring in liquid ammonia. In **Fig. 9(b)**, the $T_c$-$p$ curve of $Li_x(NH_3)_yFe_2Se_2$ is shown[74]. Applying pressure into this compound decreases the $T_c$ up to 2GPa, above which the $T_c$ increases from 22K to 56K, forming two-dome structure against the applied pressure. A similar phase diagram is also found in $Li_{1-x}Fe_xOHFe_{1-y}Se$[77,79]. This compound is also an electron-doped iron selenide system and the amount of electron doping is tuned by changing Li/Fe ratio in the blocking $Li_{1-x}Fe_xOH$ layer sandwiched by the FeSe layers. In **Fig.9(c)**, not only the $T_c$-$p$ curve of $Li_{1-x}Fe_xOHFe_{1-y}Se$ but also the change of exponent α in the resistivity ($\rho = AT^\alpha + \rho_0$) of normal state is shown as contour plot. As is observed in $LaFeAsO_{1-x}H_x$, the non-Fermi liquid state is grown above the higher $T_c$ phase (SC2).

The non-monotonic change of $T_c$ with doping or applied pressure is observed if a new phase that impedes superconductivity is developed or if more than two pairing interactions exist. The former case is seen in cuprate superconductors, $La_{2-x}Ba_xCuO_4$ (**Fig. 10(a)**)[80–85]. At $x \approx 0.125$, the superconductivity is strongly suppressed due to charge ordering (CO) and spin ordering (SO) in which the doped holes are segregated to form hole-rich stripes order in the $CuO_2$ planes. The ordering of charge and spin introduces strains into the $CuO_2$ planes, thus forming a superstructure detectable by diffraction techniques. These competition between CO or SO and superconductivity at $x = 0.125$ is called 1/8 anomaly[86]. An example of the latter case, that is, a two-dome structure derived from two paring interactions, is heavy fermion superconductor $CeCu_2(Si_{1-x}Ge_x)_2$[87–93]. **Fig. 10(b)** is a diagram of transition temperature versus a relative pressure which reflects the inverse unit cell volume. Applying pressure reduces the volume, thus acting a positive pressure, while Ge substitution for Si expands the volume which introduces a negative pressure. At the low pressure region, AFM phase is developed below $T = 2K$. The first superconducting phase (SC1) emerges in the relative pressure above −1.5GPa, while SC2 with a higher $T_c$ appears at $p > 1.2GPa$. The Fermi liquid state above the $T_c$ of SC1 is converted to the non-Fermi liquid state above the SC2 phase on applying pressure. The SC1 is believed to be mediated by the AFM spin fluctuations, while the SC2 is to be derived from a valence fluctuations of Ce. The two-dome structure merges into a single dome with a higher $T_c$ at $x = 1$ as shown in dashed line.

### 3.3. Bipartite parent phase in 1111-type

After we discovered the two-dome structure in $LaFeAsO_{1-x}H_x$, a synchrotron x-ray diffraction and inelastic neutron scattering measurements were performed on the sample with $x = 0.2$ to see whether a charge or spin ordering that destroys superconductivity develops or not as seen in $La_{2-x}Ba_xCuO_4$. However, neither structural nor magnetic transitions were observed and spin fluctuations resulting from the FS nesting was completely suppressed at $x = 0.2$[94]. On the other hand, the inelastic neutron scattering measurement on the sample with $x = 0.40$ detected strong spin fluctuations at different wave number to the FS nesting vector, even though the FS nesting should be deteriorated by the heavy electron doping[95]. This observation at the $x = 0.40$ is consistent with the non-Fermi liquid behavior, often referred to as an indicator of development of AFM spin fluctuations[69]. Therefore, in order to investigate the presence of a new spin, charge or orbital ordering in the over-doped region of LaFeAsO$_{1-x}$H$_x$, we conducted neutron diffraction, muon spin relaxation, and synchrotron x-ray diffraction measurements on the samples with $x > 0.4$. **Fig. 11(a)** shows the renewed electronic phase diagram of LaFeAsO$_{1-x}$H$_x$[26,96,97]. A tetragonal-orthorhombic transition and following PM-AFM transition are found after the SC2 phase is suppressed on doping. For the SC1 phase, the non-doped AFM phase (AFM1) with the structural transition is regarded as the parent phase. Thus, the AFM phase (AFM2) with a structural transition adhering to the SC2 should be regarded to be a 'doped' parent phase, and it would be reasonable to think that the SC2 emerges by mediating fluctuations of spins, orbital or charge derived from the doped parent phase. In $0.40 \leq x \leq 0.45$, the SC2 and AFM2 coexist macroscopically, where the volume fraction of superconductivity decreases as $x$ increases, while the magnetic volume fraction measured by muon spin relaxation increases up to more than 90% at $x = 0.45$. The $T_s$ and $T_N$ are lower than those of undoped parent phase, and the temperature gap ($T_s - T_N$) is also smaller. The bipartite parent phase is also confirmed in the high-$T_c$ and single dome system, SmFeAsO$_{1-x}$H$_x$[98]. As is observed in LaFeAsO$_{1-x}$H$_x$, the doped parent phase firstly shows a structural transition on cooling and the PM-AFM transition occurs subsequently. Contrary to the LaFeAsO$_{1-x}$H$_x$, the magnetic transition in the doped parent phase occurs by two steps: from PM to multi-$k$ AFM (C+IC-AFM2) and from the multi-$k$ AFM to incommensurate AFM (IC-AFM2), where the multi-$k$ AFM structure has commensurate and incommensurate $k$ vectors. The complex magnetic structure of AFM2 in SmFeAsO$_{1-x}$H$_x$ is originated to interactions between spins in Fe-3$d$ and Sm-4$f$ orbitals.

**Figs 12** illustrate the crystal structure of high-*T* tetragonal phase, low-*T* orthorhombic phase of non-doped and doped parent phases[31,96]. The structural transition of non-doped parent phase accompanies a rhombic distortion of the unit cell of tetragonal phase (see the blue solid line in **Fig. 12(a)** and the dashed line in **Fig. 12(b)**). In the case of structural transition at the doped parent phase, an off-centering of Fe inside the As4 tetrahedron, corresponding to anti-parallel displacement of Fe to As, takes place in addition to the rhombic distortion of *ab* plane. Such a deformation with a loss of the inversion symmetry is not common in metals, since static internal electric fields are immediately screened by conduction electrons[99,100]. This implies that electron correlations in Fe-*d*-orbitals may profoundly affect the unique structural transition.

In **Figs 13**, the AFM structures of doped parent phase are compared with those of other IBSCs. The AFM structure of non-doped LaFeAsO, SmFeAsO, and BaFe$_2$As$_2$ is a stripe-type ordering with a spin direction parallel to the propagation vector *k* (**Figs 13(a) and (b)**)[98,101–103]. In this AFM, both the direction of *k* vector (parallel to $\boldsymbol{a}_T + \boldsymbol{b}_T = \boldsymbol{a}_O$) and the length $|\boldsymbol{k}|$ (= $\sqrt{2}\pi/|\boldsymbol{a}_T|$) correspond to those of FS nesting vector connecting $\Gamma$ and M points[104–106]. The magnetic moment of each AFM structure is less than $1\mu_B$/Fe, consistent with the itinerant SDW scenario. For FeTe, however, the direction ($\boldsymbol{k} // \boldsymbol{a}_T$) and length ($|\boldsymbol{k}| = \pi/|\boldsymbol{a}_T|$) of *k* vector are totally different from the nesting vector, indicating that the AFM is nothing to do with the FS nesting (**Fig. 13(c)**)[107]. The magnetic moment of $1.9\mu_B$/Fe larger than that of non-doped iron arsenides also supports the picture of localized magnetism. For the structures of AFM2 phase in LaFeAsO$_{0.49}$H$_{0.51}$ and SmFeAsO$_{0.27}$H$_{0.73}$, the size of magnetic moment on Fe is more than twice as large as those of non-doped parent phase. For the AFM of LaFeAsO$_{0.49}$H$_{0.51}$, the *k* vector is the same as that of LaFeAsO even though the heavy electron doping largely degrades the FS nesting. The Hall effect measurements on AFM of 1111-and 122-type detect a large reduction of carrier concentration below $T_N$ owing to the opening of the SDW gap, whereas such a change in carrier concentration is not observed for the AFM in LaFeAsO$_{0.49}$H$_{0.51}$ and FeTe[108–112]. These data also suggest that a more localized picture rather than the itinerant SDW is appropriate for the AFM of LaFeAsO$_{0.49}$H$_{0.51}$ and FeTe[96,113–115].

### 3.4. Electronic structure

The emergence of localized magnetism in the heavily electron doped region is unexpected, because electronic correlations are widely supposed to be weaken by heavy doping. In order to examine effects of electron doping and the origin of the doped parent phase, DFT calculations are performed for the SmFeAsO$_{1-x}$H$_x$. Here, the doping effect is simulated by a virtual crystal approximation. **Fig. 14(a)** shows the $x$ dependence of orbital energy, $\varepsilon$, of Fe-3$d$ and As-4$p$. These energies are aligned by the Sm 5$s$ core level (−41- −40eV below their $E_F$). As electrons are doped into Fe-3$d$ orbitals, both the $\varepsilon$ of Fe-3$d$ and As-4$p$ orbitals gradually increases from the Sm's core level, suggesting that these energy levels becomes shallower with respect to a vacuum level. The upshift in the $\varepsilon$ of Fe-3$d$ can be understood as a consequence of Fe's valence change from 2+ to ~1+. Generally speaking, the energy level of Fe-3$d$ orbital strongly depends on the valence state of Fe. For instance, a highly oxidized Fe$^{4+}$ state is as deep as the energy level of O-2$p$, making a strong covalent bonding with oxygen, while the Fe-3$d$ level becomes much shallower as the valence number is reduced, yielding a more ionic Fe-O bonding (see **Fig. 14(b)**)[116,117]. The $\varepsilon$ of As-4$p$ orbitals is also slightly influenced by the doping, probably because the Madelung energy around the $As^{3-}$ is decreased due to the reduction of valence state of Fe. As a result, the shallowing of Fe-3$d$ energy level makes the Fe-As bond more ionic, in other word, the energy gap between ε of Fe-3$d$ and As-4$p$ orbitals is enlarged. In **Fig. 14(c)** the ε of Fe and As are compared with those of other IBSCs. The ionicity between Fe-3$d$ and anion's $p$ orbitals systematically increases from phosphide (LaFePO) to arsenides (1111, 122, and 111-types) and chalcogenides except the heavily electron doped SmFeAsO (SmFeAsO$_{0.18}$H$_{0.82}$). From this trend, we can say that the ionicity of Fe-As bond in SmFeAsO$_{0.18}$H$_{0.82}$ is close to that of the most ionic Fe-Se bond. The ionicity is also known as an important measure of electron correlations in IBSCs[118]. **Fig. 14(d)** shows onsite Coulomb ($U$) and exchange ($J$) interactions of iron chalcogenides, arsenides and phosphide. Both values increases as the ionic character is enhanced. The larger effective interaction in the ionic IBSCs is attributed to a poor screening of Coulomb interactions in Fe by anion $p$ orbitals. Considering this trend between the ionicity and the strength of electron correlations, electrons in heavily electron-doped SmFeAsO can be regarded to be correlated as strong as iron chalcogenides.

The enhanced ionic character in Fe-anion bonding also influences on the band width. **Fig. 15(a)** is the $x$ dependence of nearest neighbor hopping parameter ($t_1$) and next nearest neighbor hopping parameter ($t_2$) of Fe-3$d_{xy}$ orbital in SmFeAsO$_{1-x}$H$_x$[98,119]. The Fe-3$d_{xy}$ orbital is important to examine the electronic structure of IBSCs since it is the one of the main contributors to the bands across $E_F$[120]. The $t_1$ of Fe-3$d_{xy}$ decreases rapidly with doping and approaches zero at $x \sim 0.5$ above which the sign is converted. The reduction of $t_1$ is due to a destructive interference of two components with an opposite sign: one is a direct hopping between nearest neighboring Fe, and the second is indirect hopping through the anion[119,121–123]. This situation is illustrated in **Fig. 15(b)**. The electron doping enlarges the spread of Fe-3$d$ and As-4$p$ orbitals and enhances both direct and indirect $t_1$. However, because the energy gap between Fe-3$d_{xy}$ and As-4$p$ increases due to the enhanced ionicity, the increase of indirect $t_1$ is limited. Consequently, the negative direct $t_1$ almost exactly cancels the positive indirect $t_1$, resulting in vanishing effective $t_1$ of Fe-3$d_{xy}$. Z. P. Yin *et al.*, summarizes the $t_1$ and $t_2$ of a large number of IBSCs and compared those with the physical properties such as the ordered magnetic moments, effective masses and fluctuating PM moments based on DFT calculations[122]. The results are shown in **Fig. 15(c) and (d)**. The $t_1$ is reduced and the magnitude of $t_1$ and $t_2$ is reversed by changing the character of Fe-anion bond from covalent to ionic, in which the ordered and fluctuating magnetic moments are enhanced. The authors ascribe the reduction of $t_1$ to a long Fe-anion bonding, that is, a structural effect. However, for SmFeAsO$_{1-x}$H$_x$, the electron doping effect rather than the structural effect primarily affects the $t_1$ reduction[98].

So, the primal question is what kind of band structure derived from Fe-3$d_{xy}$ orbital bears the strong electron correlation in SmFeAsO$_{1-x}$H$_x$? **Fig. 16(a)** shows the band structure of SmFeAsO$_{0.53}$H$_{0.47}$ with the $t_1$ of 3$d_{xy}$ ≈ 0. There are totally 10 bands in the energy range from −2 to 2eV, corresponding to 3$d$ orbitals of two Fe being inside the unit cell. Here, the two bands derived from 3$d_{xy}$ are emphasized as a contour plot. Since the $t_1$ of 3$d_{xy}$ is almost zero, the band structure of 3$d_{xy}$ orbital can be simply simulated by the tight binding model: $E(k) = \varepsilon + 2t\,[\cos(k_x) + \cos(k_y)]$, where $t$ corresponds to $t_2$ of 3$d_{xy}$. The result is shown as solid green lines. The discrepancy at X point and the M-Γ line is due to a hybridization of 3$d_{xy}$ and 3$d_{yz/zx}$.

The dispersion of Fe-$3d_{xy}$ bands with only $t_2$ bear a close resemblance to that of Cu-$3d_{X2-Y2}$ band of non-doped cuprates. **Fig. 16(b)** shows the band structure of $HgBa_2CuO_4$ in which the contribution from Cu-$3d_{X2-Y2}$ is shown by the contour plot. It is widely known that the dispersion of Cu-$3d_{X2-Y2}$ band across $E_F$ is also described well by the same tight binding model $E(k) = \varepsilon + 2t\ [\cos(k_x) + \cos(k_y)]$ using $t = t_1$ between neighboring Cu-$3d_{X2-Y2}$ orbitals. The result is shown as a solid green line.

The resemblance of band structures in the heavily electron doped SmFeAsO and the non-doped cuprate is understood by writing down the corresponding orbital configurations of Fe-$3d_{xy}$ and Cu-$3d_{X2-Y2}$. In **Fig. 16(c)** an orbital configuration of Fe-$3d_{xy}$ in $SmFeAsO_{0.53}H_{0.47}$ with $t_1 \approx 0$ is illustrated. There are two Fe and two As in the tetragonal unit cell. Here the nearest neighbor hopping ($t_1$) corresponds to a hopping between the two Fe in a same unit cell, while the next nearest neighbor hopping ($t_2$) does to a hopping into the Fe in neighboring unit cell. The $t_2$ which is dominant in this system is mediated by the As-$4p$ orbital and the hopping distance is identical to the lattice parameter $a$ of ~4Å. In the case of Cu-$3d_{X2-Y2}$ in non-doped cuprate (**Fig. 16(d)**), the dispersion of the single band across $E_F$ is primarily determined by the nearest neighbor hopping ($t_1$) of Cu-$3d_{X2-Y2}$, which is mediated by oxygen $2p$ and the corresponding hopping distance corresponds to the lattice parameter $a$ of ~4Å[124]. Note that the pathway of $t_2$ in Fe-$3d_{xy}$ and $t_1$ in Cu-$3d_{X2-Y2}$ are almost identical except the position of anion: As is located above and below the Fe plane while O is in the Cu plane. This is the reason why the band structures of heavily electron doped SmFeAsO and the non-doped cuprate are similar to each other.

To date, some mechanisms for the emergence of doped parent phase have been proposed. Yamakawa *et al*. proposed spin and orbital orderings with the incommensurate Fermi surface nesting at LaFeAsO$_{1-x}$H$_x$ with $x = 0.4$[125]. Suzuki *et al*. indicated that the next nearest neighbor hoppings ($t_2$) between Fe-3$d_{xy}$ orbitals dominate over the nearest neighbor ones ($t_1$), and the resulting $J_1 < J_2$ ($J_i \approx t_i^2/U$) is naively expected to induce the stripe-type ($\pi$, 0) spin fluctuations[119]. Moon *et al*. examined the electronic state of LaFeAsO$_{1-x}$H$_x$ by dynamical mean-field theory and attributed the emergence of the doped parent phase into the elongation of Fe-As which enhances the local correlation for the 3$d_{xy}$ orbital via the reduction of $t_1$[126].

In multi-orbital systems such as IBSCs, the electron localization by onsite Coulomb interactions $U$ is less significant, because electrons easily move to other degenerated orbitals. However, if this inter-orbital transfer of electron is suppressed by a strong Hund coupling, electrons tends to localize and an energy gap is open in a specific orbital with a moderately large $U$[127]. This is the idea of orbital-selective Mott transition (OSMT), that is, a part of the conduction electrons is localized while the rest remains itinerant[128]. The large $U$ expected in the Fe-3$d_{xy}$ orbital of heavily electron-doped IBSCs and the similarity of band structure between heavily electron-doped Fe-3$d_{xy}$ in IBSCs and Cu-3$d_{X2-Y2}$ in cuprates are reminiscent of the possibility of OSMT. Such an orbital selective motion of electrons may also help to understand the unique structural transition accompanying with the off-centering of Fe in the doped parent phase of 1111-type.

## 4. Perspective

A hot issue in unconventional superconductors is why the 2nd highest-$T_c$ of 56 K next to cuprates is accomplished in the electron-doped 1111-type iron-based superconductors. The hydride ion substitution for the oxygen site enabled a high concentration electron doping in the 1111-type and unveiled the unique properties such as the two-superconducting-dome structure, the conversion of the low-$T_c$ two domes to high-$T_c$ single dome by the structural tuning, and the bipartite parent phase. Considering the fact that the single parent systems such as 111 and 122-type exhibit relatively low $T_c$ ~ 20-30K, a cooperative enhancement of spin[33,34], orbital[129,130] or charge[131] fluctuations derived from the bipartite parent phase may be a key to achieving high-$T_c$ superconductivity in IBSCs. For the origin of doped parent phase, we discussed the importance of increase in ionicity of Fe-anion bonding enhanced by electron doping. In this regards, the high $T_c$ two-dome structure in electron-doped iron chalcogenides observed under high pressure may provide a clue to achieve higher $T_c$ over 56K (**Fig. 9(b) and (c)**)[74,77]. For the 1111-type, the shape of superconducting dome and $T_c$ were sensitive to both the carrier doping and the local structure around Fe. Further doping or elemental substitution might merge the two superconducting domes into a single dome with a much higher $T_c$. In addition, it may be interesting to explore another parent phase under higher pressure. In fact, the non-Fermi liquid behavior and a convex-shaped $\rho$-$T$ curve at the high pressure region seem to suggest that the electrons are on the verge of forming a new phase such as AFM[74,77,109]. The hydride ion doping may also be an attractive way to explore new superconductors. Recently, LaFeSiH is reported to become superconducting below $T_c$ of 11K by inserting hydrogens into an interstitial site of LaFeSi[132]. In this case, the hydrogens accept electrons from La and/or Fe and become hydride ions.

Last, we would like to address the critical role of hydride ion in functional oxide materials by introducing two examples. First is transparent amorphous oxide semiconductors with a large electron mobility. A representative material is In-Ga-Zn-O (IGZO)[133,134] which is now used as the channel layers of thin film transistors (TFTs) to drive pixels of advanced flat panel displays[135]. Recently, a large amount ($\sim 10^{19\text{-}20}$cm$^{-3}$) of hydride ion bonded with the metal cations was found in IGZO thin films deposited by a conventional sputtering method which are applied for practical productions. This hydride ion is much more stable thermally than protons incorporated in the form of hydroxy group in the thin films and gives rise to energy levels above the valence band maximum which are closely related with a TFT instability under a negative bias and illumination stresses[136]. Another is active roles of hydride ion in catalysts for ammonia synthesis at mild conditions. It was clarified that a reversible exchange between hydride ion and anionic electron is a key to enhancement of catalytic activity at low temperatures[137]. Hydrogen is the most abundant element in universe and is the simplest atom. Although most oxide materials incorporate or contaminate hydrogen species, their critical role in the functionality of the material has only recently been microscopically understood. Careful characterization of materials using an appropriate technique and a combination with computations by taking into account the bipolarity of hydrogen is highly required[138,139]. We expect that the utilization of unique nature of H$^-$ ion in solids would extend the possibility of design for various materials including superconductors.

## Acknowledgement

We thank Dr S. Matsuishi of Tokyo Institute of Technology for discussions. This study was supported by MEXT Elements Strategy Initiative to form the Core Research Center. A portion of this work was supported by JSPS KAKENHI Grant Numbers 17H06153, JP18H05513, 18K13499 and the PRESTO program (No.JPMJPR19T1) of the JST.

# Figures and figure legends

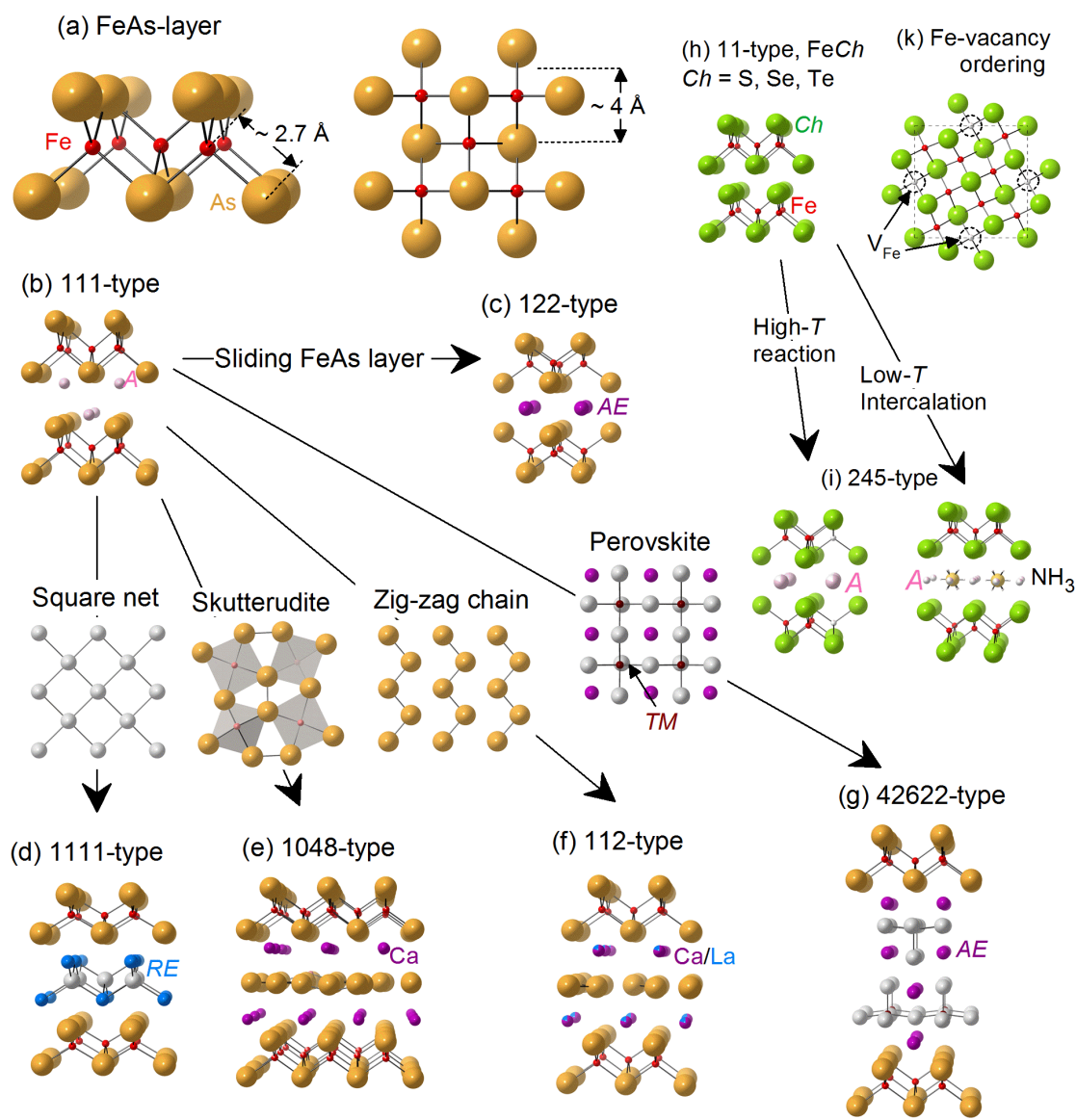

Fig.1 (color online) Crystal structures of IBSCs.

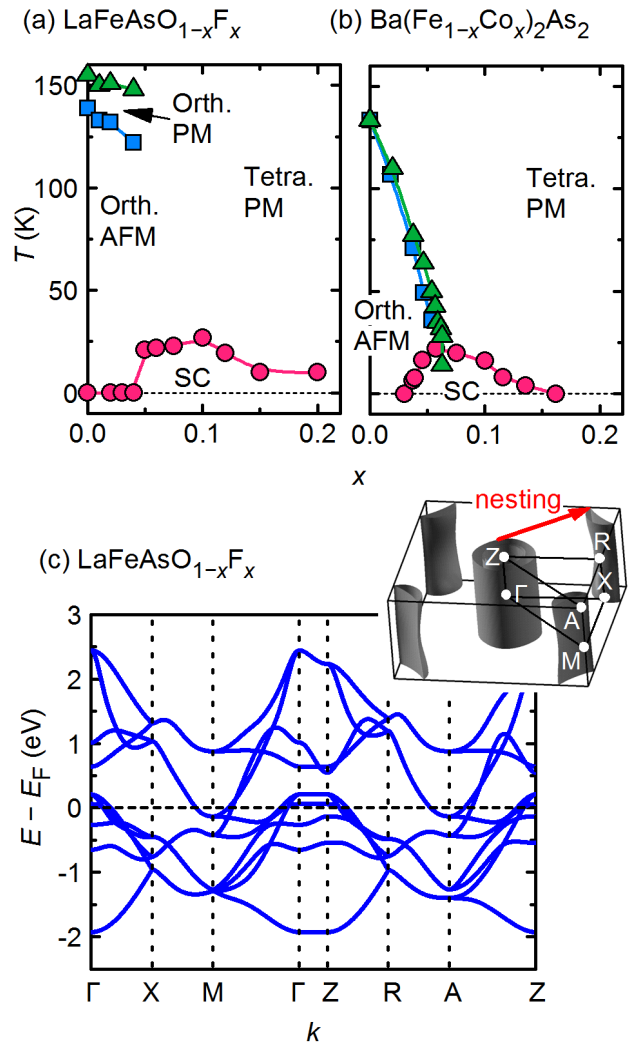

Fig.2 (color online) (a, b) Electronic phase diagrams of LaFeAsO$_{1-x}$F$_x$[9,33,3] (a) and Ba(Fe$_{1-x}$Co$_x$)$_2$As$_2$[28,29] (b). (c) Band structure of high-temperature tetragonal LaFeAsO. In the inset, the Fermi surface is shown. The red arrow represents the nesting vector connecting the hole and electron pockets at Γ and M points, respectively.

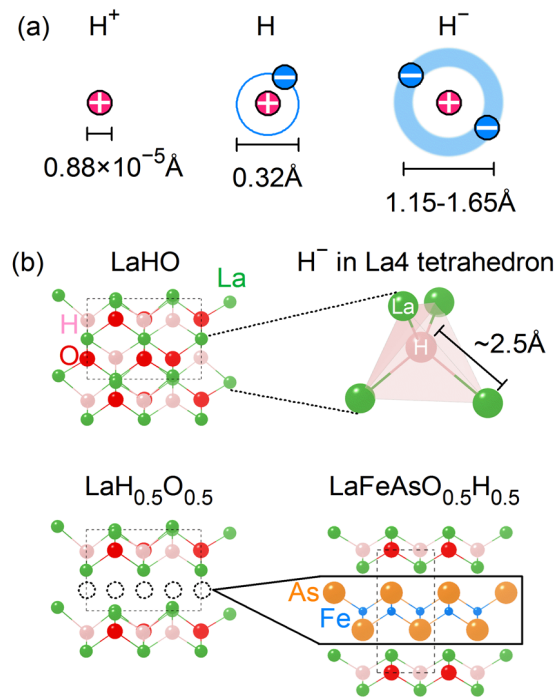

Fig.3 (color online) (a) Schematic of electron configuration of $H^+$, $H^0$, and $H^-$. The root-mean-square charge radius of $H^+$ ($0.88\times10^{-5}$Å)[49], covalent radius of $H^0$ (0.32 Å)[50] and the ionic radius of $H^-$ (1.15-1.65Å)[48] are also shown. (b) Crystal structures of LaHO, LaHO with an ordered H and O vacancy, and $LaFeAsO_{0.5}H_{0.5}$.

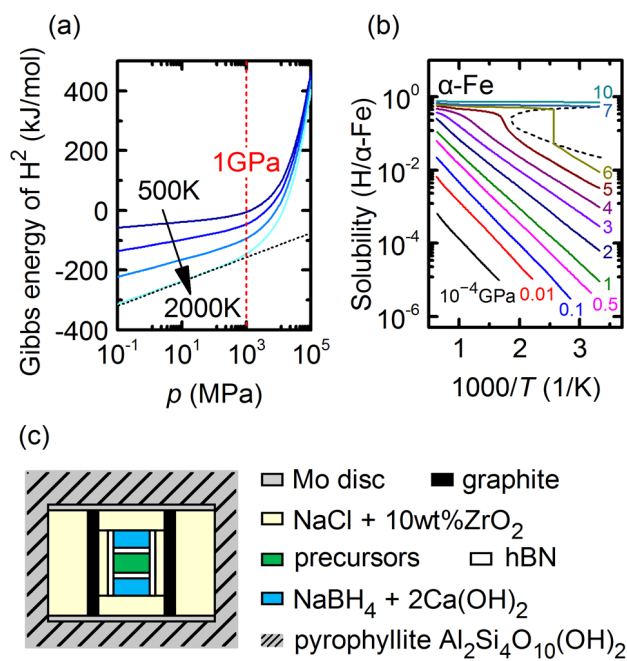

Fig. 4 (a) (color online) Pressure dependence of Gibbs energy of gaseous $H_2$[61]. The solid lines correspond the Gibbs energy at 500, 1000, 1500, and 2000K from top to bottom. Black dotted line represents the data calculated assuming ideal gas. (b) Hydrogen solubility versus inverse temperature curves at different pressures for α-Fe. Black dashed line represents the boundary of the two hydride phases region (spinodal decomposition). (c) Sample cell assembly for high pressure synthesis[45,63–65].

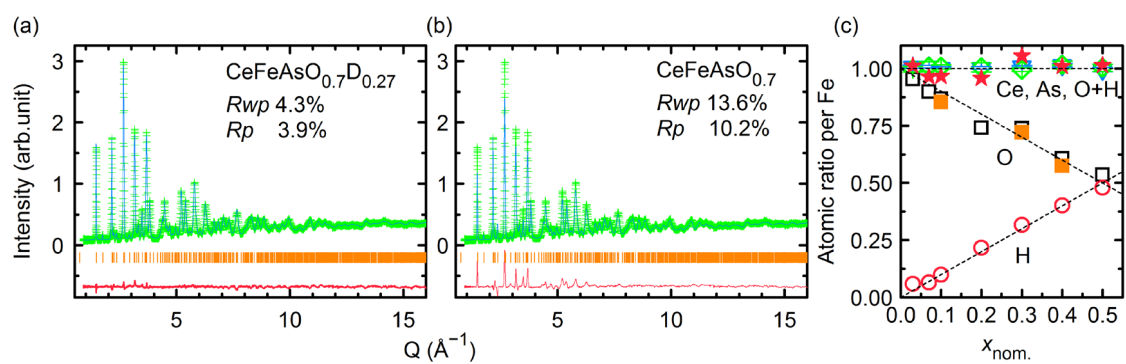

Fig.5 (color online) (a, b) Neutron powder diffraction patterns and simulated patterns[46]. The deuterium-substitution and oxygen-vacancy models are used for (a) and (b), respectively. (c) The atomic ratio per Fe as a function of nominal *x*. The amounts of Ce, As, and the summation of O and H are shown as blue triangles, green rhombuses, and red stars. The hydrogen amounts shown as red open circles were obtained by thermal desorption gas spectroscopy, and oxygen amounts were measured by NPD (orange filled squares) and electron probe micro analyzer (black open squares).

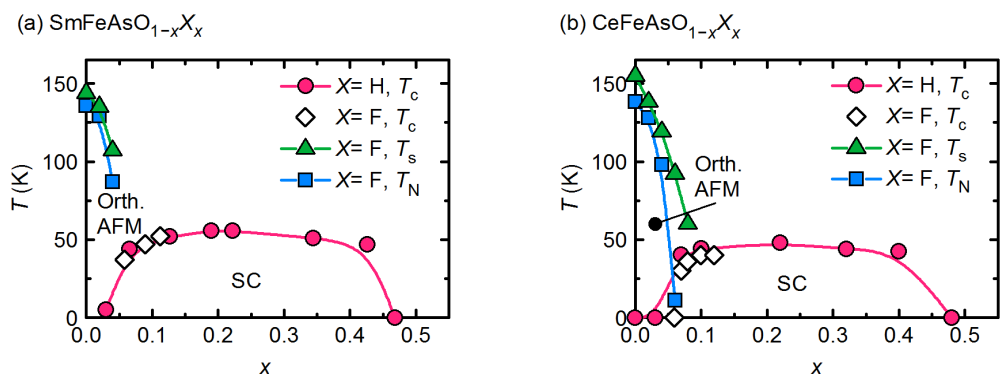

Fig.6 (color online) (a, b) Electronic phase diagrams of SmFeAsO$_{1-x}$(H, F)$_x$[45] and CeFeAsO$_{1-x}$(H, F)$_x$[71].

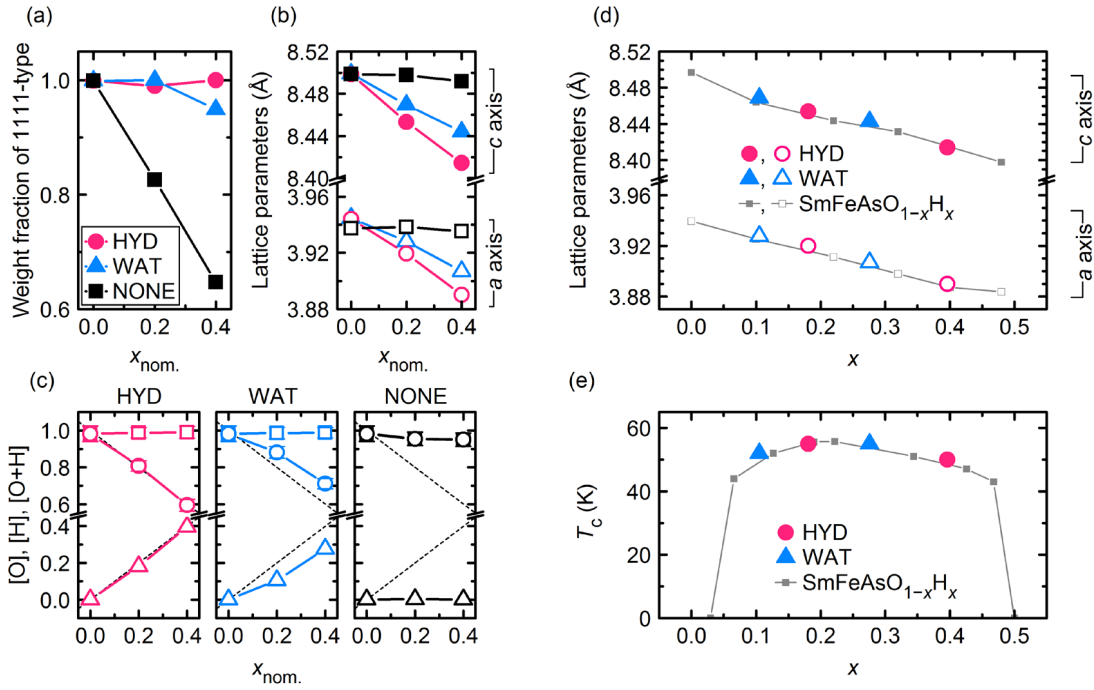

Fig.7 (color online) (a) $x_{nom.}$ dependence of weight fraction of 1111-type phase[67]. Here, HYD, WAT, and NONE represent samples synthesized under atmospheres of hydrogen gas, water, and arogon without $H_2$ or $H_2O$, respectively. (b) $x_{nom.}$ dependence of lattice parameters of 1111-type phase. (c) The amount of oxygen [O], hydrogen [H] and the summation [O+H] in 1111-type as a function of $x_{nom.}$. The dashed lines are [H] = $x_{nom.}$, [O] = $x_{nom.}$, and [O+H] =1 for a guide of eyes. (d) Lattice parameters of HYD, WAT samples and $SmFeAsO_{1-x}H_x$. (e) $T_c$ of HYD, WAT samples and $SmFeAsO_{1-x}H_x$.

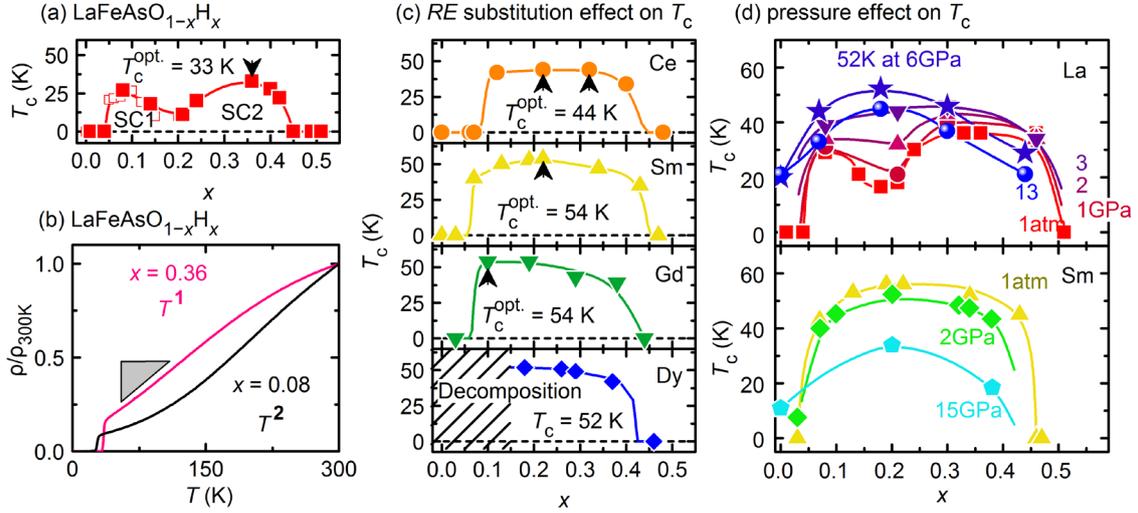

Fig.8 (color online) (a) $T_c$-$x$ curves of LaFeAsO$_{1-x}$H$_x$[47] (filled squares) and LaFeAsO$_{1-x}$F$_x$[27] (open squares). The arrow indicates an optimum $x$. (b) Comparison of $\rho$-$T$ curves at $x$ = 0.08 and 0.36 of LaFeAsO$_{1-x}$H$_x$[27]. (c) Comparison of $T_c$-$x$ curves of $RE$FeAsO$_{1-x}$H$_x$ with $RE$ = Ce, Sm, Gd, and Dy[27]. In Dy-1111, no 1111-type phase was obtained at $x$ < 0.18. (d) Pressure dependence of $T_c$-$x$ curves of LaFeAsO$_{1-x}$H$_x$ (upper panel) and SmFeAsO$_{1-x}$H$_x$ (lower panel)[70].

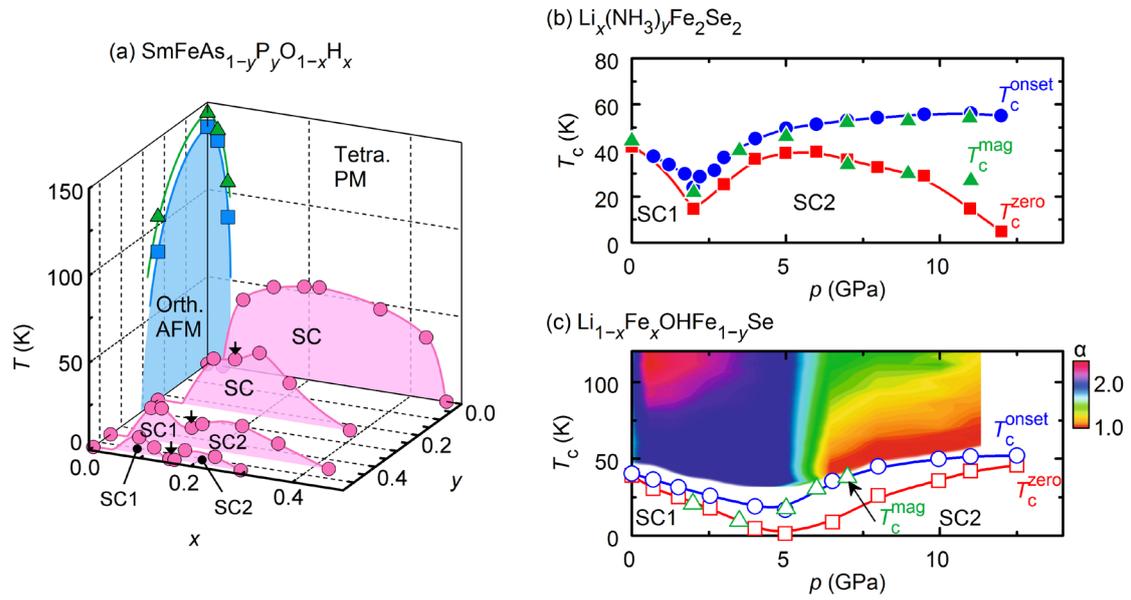

Fig.9 (color online) (a) Phase diagram of SmFeAs$_{1-y}$P$_y$O$_{1-x}$H$_x$[71]. (b) $T_c$-$x$ curves of Li$_x$(NH$_3$)$_y$Fe$_2$Se$_2$[74]. $T_c^{onset}$, $T_c^{zero}$, and $T_c^{mag}$ are the onset temperature of resistivity drop, the zero resistivity temperature, and $T_c$ determined by susceptibility measurements. (c) $T_c$-$x$ curves of Li$_{1-x}$Fe$_x$OHFe$_{1-y}$Se[77]. The contour plot above those $T_c$ represents a magnitude of exponent $\alpha$ in $\rho = \rho_0 + AT^\alpha$ equation.

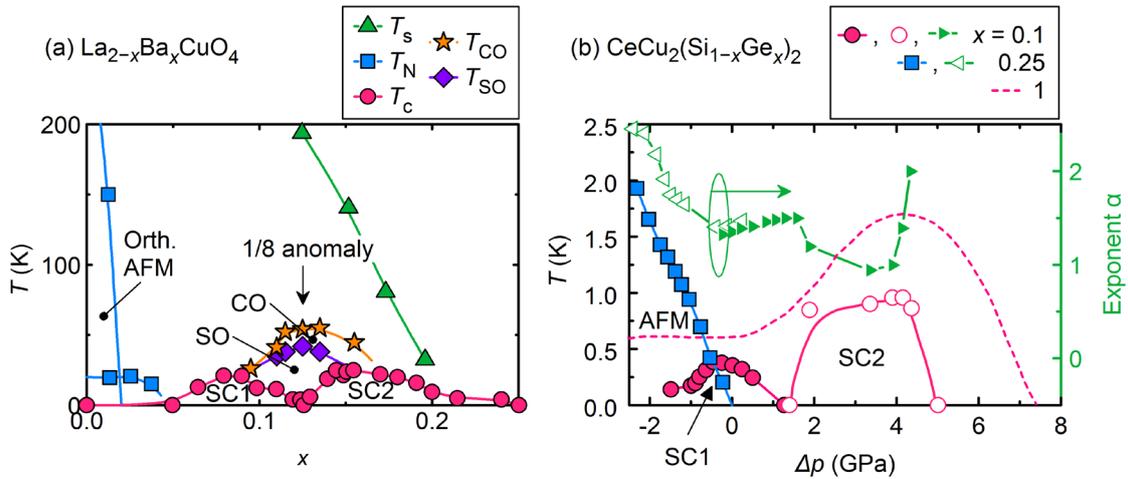

Fig.10 (color online) (a) Phase diagram of La$_{2-x}$Ba$_x$CuO$_4$[80–85]. $T_s$, $T_N$, $T_{CO}$, and $T_{SO}$ respectively represent transition temperatures of crystal structure from tetragonal (*I*4/*mmm*) to orthorhombic (*Cmca*), magnetic structure from PM to checkerboard-type AFM, charge ordering (CO) from orthorhombic (*Cmca*) to tetragonal (*P*42/*ncm*), and spin ordering (SO) from PM to stripe-type AFM. (b) Phase diagram of CeCu$_2$(Si$_{1-x}$Ge$_x$)$_2$[87–93]. The horizontal axis is relative pressure $\Delta p$ reflects the inverse unit cell volume. Blue squares, red circles, and green open triangles are taken from the samples with $x = 0.1$, while red open circles and green filled triangles are from $x = 0.25$. The dash line denotes $T_c$ at $x = 1$. The green symbols represent a magnitude of exponent $\alpha$ in $\rho = \rho_0 + AT^\alpha$ equation.

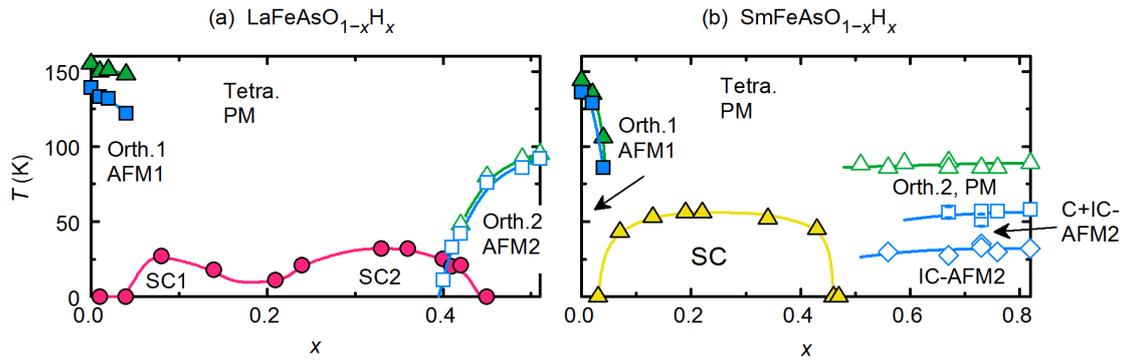

Fig.11 (color online) (a, b) Renewed phase diagram of LaFeAsO$_{1-x}$H$_x$[26,47,96] (a) and SmFeAsO$_{1-x}$H$_x$[98] (b). C-AFM and IC-AFM in (b) represent a commensurate and incommensurate AFM, respectively. C+IC AFM means a multi-$k$ AFM structure with the commensurate and incommensurate propagation vectors ($k$ vector).

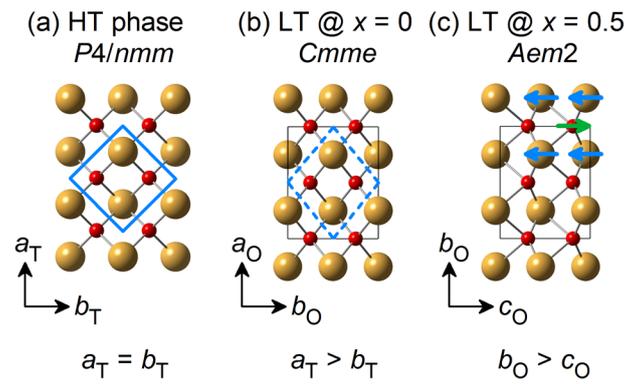

Fig.12 (color online) (a-c) Crystal structures of LaFeAsO$_{1-x}$H$_x$ at high temperature (a), low temperature at $x = 0$[30,31] (b) and $x = 0.5$[96] (c). Blue solid line in (a) is the tetragonal unit cell, and blue dashed line in (b) corresponds to the tetragonal cell. In the orthorhombic cell at $x = 0.5$ shown in (c), the crystallographic axes are changed; the tetragonal $a$- and $b$-axes ($a_T$, $b_T$) changes to the orthorhombic $b$- and $c$-axis ($b_O$, $c_O$), respectively. Green and blue arrows respectively denote displacements of Fe and As in the tetragonal-orthorhombic transition.

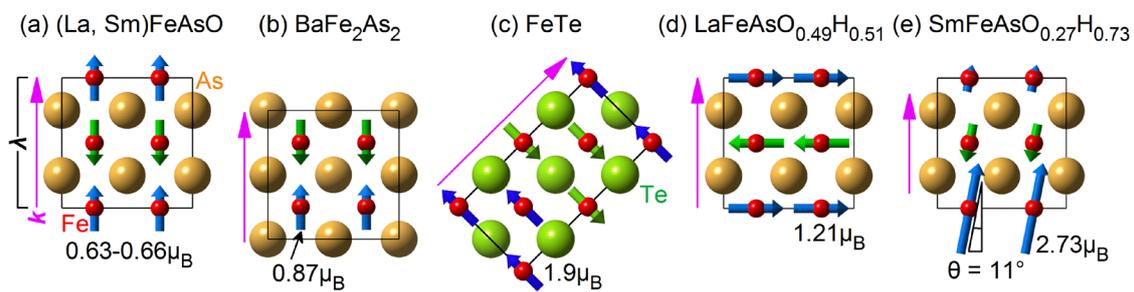

Fig.13 (color online) (a-e) AFM structures of (La, Sm)FeAsO[98,101] (a), BaFe$_2$As$_2$[102,103] (b), FeTe[114] (c), LaFeAsO$_{0.49}$H$_{0.51}$[96] (d), SmFeAsO$_{0.27}$H$_{0.73}$[98] (e).

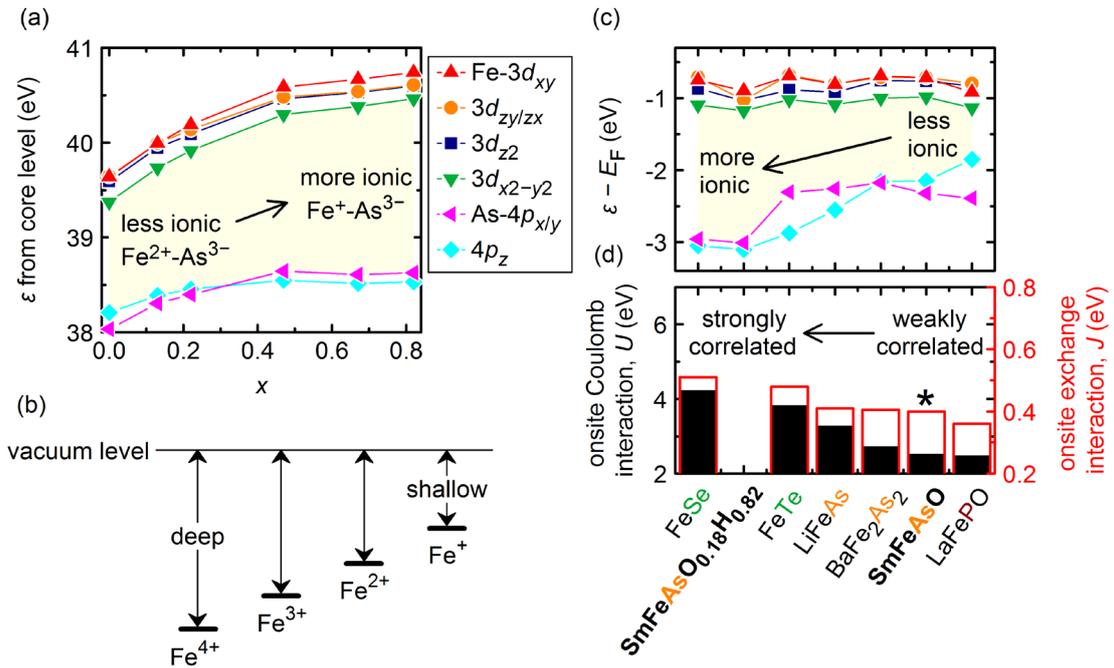

Fig.14 (color online) (a) Orbital energies, ε, of Fe-3$d$ and As-4$p$ orbitals aligned from Sm-5$s$ level as a function of $x$ in SmFeAsO$_{1-x}$H$_x$. These values correspond to a diagonal part of Hamiltonian in the Wannier basis calculated by fitting DFT band structure to Wannier orbitals[140]. For the structural model, experimental structural parameters are used, and the $f$-electrons of Sm are kept frozen in the core[140–143]fation of fo noit. (b) Schematic of energy alignment of Fe-3$d$ orbital with different valence states. (c) ε of Fe-3$d$ and anion's $p$ orbitals in various IBSCs. (d) The onsite Coulomb interaction and exchange interaction in various IBSCs[118].

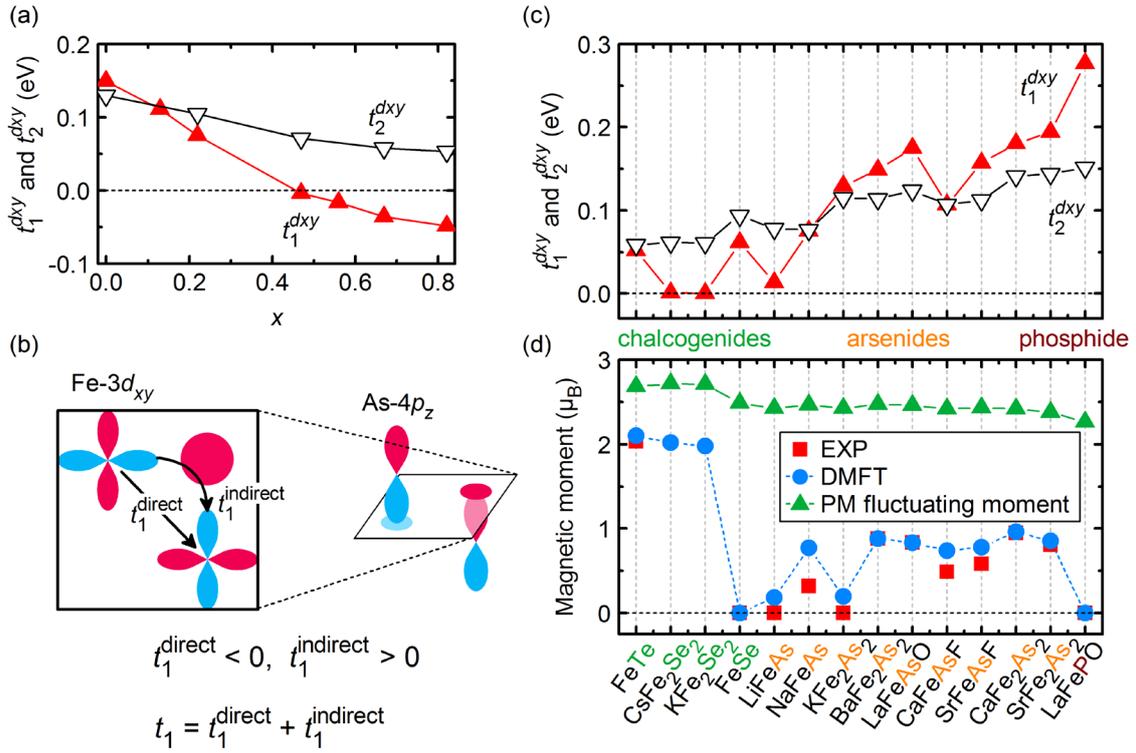

Fig.15 (color online) (a) $x$ dependence of nearest ($t_1$) and next nearest neighbor ($t_2$) hopping parameters of Fe-$3d_{xy}$ in SmFeAsO$_{1-x}$H$_x$[98,119]. (b) Schematic of the direct $t_1$ ($t_1^{\text{direct}}$) and indirect $t_1$ ($t_1^{\text{indirect}}$) of Fe-$3d_{xy}$ orbital in FeAs layer. $t_1^{\text{direct}}$ represents a direct hopping between two Fe-$3d_{xy}$, while $t_1^{\text{indirect}}$ is a hopping mainly via As-$4p_z$. (c, d) $t_1$, $t_2$, ordered magnetic moments, and paramagnetic fluctuating moment of various IBSCs[122].

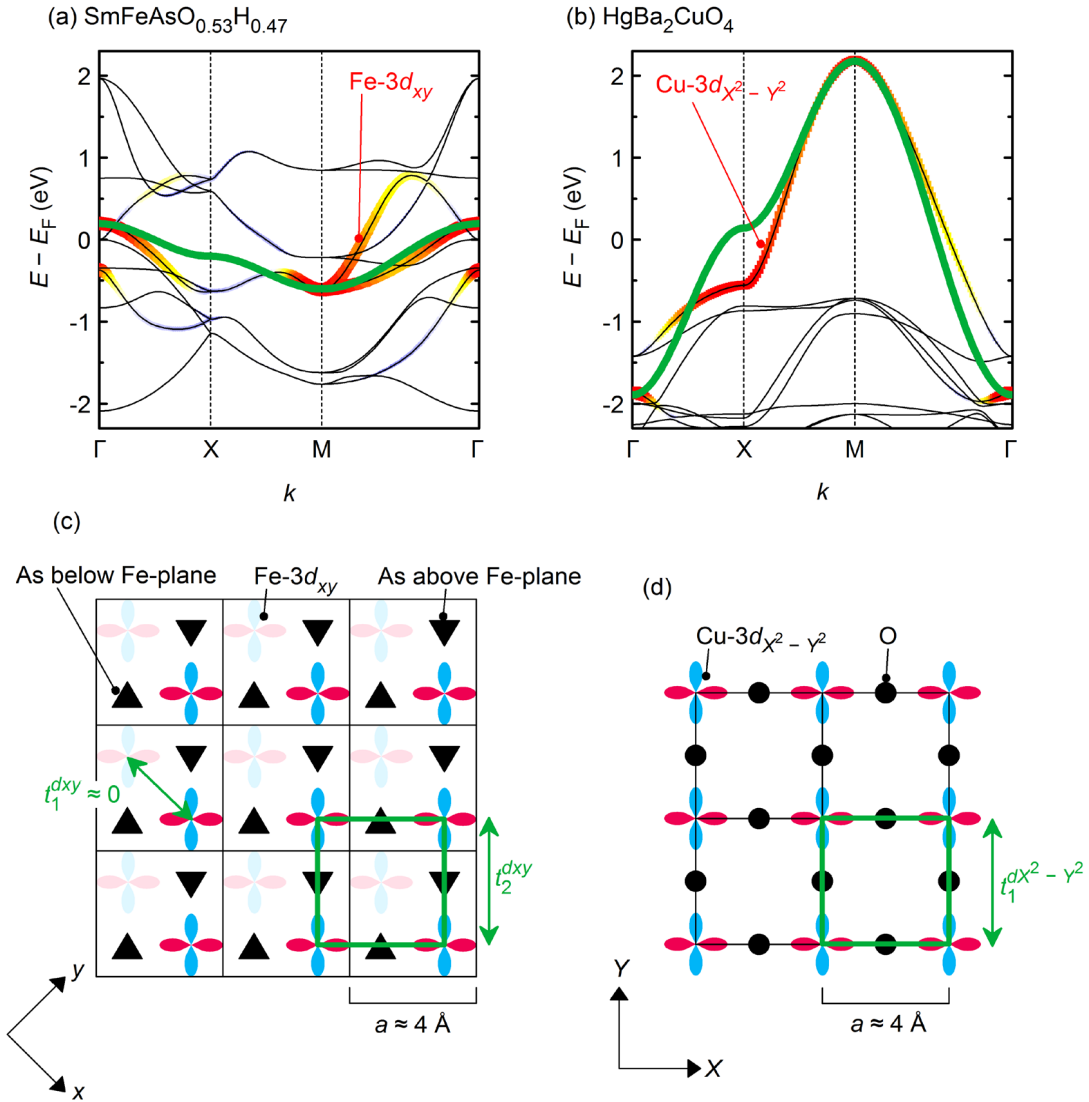

Fig.16 (color online) (a, b) Band structures of SmFeAsO$_{0.53}$H$_{0.47}$ (a) and HgBa$_2$CuO$_4$ (b). The contribution from Fe-3$d_{xy}$ and Cu-3$d_{X2-Y2}$ are emphasized as the contour plot in the band structure of SmFeAsO$_{0.53}$H$_{0.47}$ and HgBa$_2$CuO$_4$, respectively. Green solid lines denote the tight binding band structure calculated using $t_2$ of Fe-3$d_{xy}$ or $t_1$ of Cu-3$d_{X2-Y2}$. The tight binding band in SmFeAsO$_{0.53}$H$_{0.47}$ are doubly degenerated, since there are two Fe ions in the unit cell. The concave and convex shapes of the tight binding band structure in Fe-3$d_{xy}$ and Cu-3$d_{X2-Y2}$ reflect the positive sign of $t_2$ in Fe-3$d_{xy}$ and negative sign of $t_1$ in Cu-3$d_{X2-Y2}$, respectively. (c, d) The orbital configurations of Fe-3$d_{xy}$ in SmFeAsO$_{0.53}$H$_{0.47}$ (c) and Cu-3$d_{X2-Y2}$ in HgBa$_2$CuO$_4$ (d).